\documentclass[a4wide,reqno]{amsart}
\usepackage{xcolor}
\usepackage{graphicx}

\newtheorem{definition}{Definition}[section]
\newtheorem{theorem}{Theorem}%

\def \C {\mathbb{C}}
\def \D {\mathcal{D}}
\def \Hil {\mathcal{H}}

\def\h {\mathfrak{h}}
\def \R {\mathbb{R}}
\def\rmi{\mathrm{i}}
\def\rmd{\mathrm{d}}
\def\Tr{\mathrm{Tr}}
\def\u {\mathfrak{u}}

\newcommand{\norm}[1]{\left\Vert#1\right\Vert}

\begin{document}


\title{A general framework for quantum splines}

\author{L Abrunheiro}
\address{CIDMA -- Center for Research and Development in Mathematics
  and Applications, and Higher Institute of Accounting and
  Administration,  University of Aveiro, 3810-500 Aveiro, Portugal \\
}

\author{M Camarinha}
\address{Centre for Mathematics of University of Coimbra, Department
  of Mathematics, University of Coimbra, 3001-454 Coimbra, Portugal\\
}

\author{J Clemente-Gallardo}
\address{Department of Theoretical Physics,  Institute for
  Biocomputation and Physics of Complex Systems (BIFI), University of
  Zaragoza,  Edificio I+D, Mariano Esquillor s/n, 50018 Zaragoza,
  Spain\\
} 

\author{J. C. Cuch\'\i}
\address{Department of Agricultural and Forest Engineering, University
  of Lleida, Av. Alcalde Rovira Roure 191, 25198 Lleida, Spain\\
}

\author{P. Santos}
\address{Centre for Mathematics of University of Coimbra, Department of
Mathematics, University of Coimbra, 3001-454 Coimbra, Portugal
\\
and
\\
  Polytechnic Institute of Coimbra, ISEC, Department of Physics
  and Mathematics, Rua Pedro Nunes - Quinta da Nora, 3030-199 Coimbra,
  Portugal\\
}

\maketitle

\begin{abstract}
Quantum splines are curves in a Hilbert space or, equivalently, in the
corresponding Hilbert projective space, which generalize the notion of
Riemannian cubic splines to the quantum domain. In this paper, we
present a generalization of this concept to general density matrices
with a Hamiltonian approach and 
using a geometrical formulation  of quantum mechanics. 
Our main goal  is to formulate an optimal control problem for a nonlinear system  on
$\mathfrak{u}^*(n)$ which corresponds to the variational problem of
quantum splines.  The corresponding Hamiltonian equations and
interpolation conditions are derived. The results are 
illustrated with some examples and {\color{black} the corresponding quantum splines
are computed with the implementation of a suitable iterative algorithm.}

\end{abstract}


\section{Introduction}
\label{Section1}

 Quantum control theory,  and in particular quantum optimal
control, is drawing considerable attention from quite
different communities in different areas of  physics, chemistry,
applied mathematics and quantum information (see
  \cite{Werschnik2007} for a nice introduction to the area with
  several practical physical and chemical applications). This subject represents
an essential  ingredient to  use  the new quantum technologies which
are being created in these fields, with a wide range of
  applications (see, for instance,
  \cite{Bartana1993,Bartana1997,Grifoni1998,
  Khaneja2001,Ohtsuki1999,Weaver2000,Zhu1998,Zhu1998b}). 
An important problem is the design of efficient 
algorithms to generate accurately specified quantum states under
certain conditions (in minimal time, with a minimal energy investment,
etc). These algorithms usually consist in defining a suitable time
dependent Hamiltonian which drives the system to the desired quantum
states under the required conditions.

In a recent paper
\cite{Brody2012},  Brody and co-workers introduced 
the notion of quantum spline as the generalization to the quantum
domain of the familiar notion of geometric splines (see
\cite{Colombo2014,Crouch1995,Gay-Balmaz2012,Noakes1989}). {}A 
quantum spline is  the solution of the following quantum optimization
problem.  Let us consider a set of states $\{ |\psi_{0}\rangle,
\cdots, |\psi_{N}\rangle\}$ in a certain finite dimensional Hilbert
space ${\mathcal H}$ and a corresponding set of times $\{t_{0}, \cdots,
t_{N}\}$. The problem is to find a \mbox{time-dependent} Hamiltonian
$H(t)$ defining an unitary evolution which, at the given times
$t_{k}$,  passes arbitrarily close to the given states
$|\psi_k\rangle$, and such that the total change in the Hamiltonian
along the curve is minimized.  We can write it as a
variational problem asking to minimize the cost functional 
\begin{equation}   \label{eq:JproblemA}
  {\mathcal J}=\int_{t_{0}}^{t_{N}}\langle \rmi\dot H(t) \mid \rmi\dot
  H(t)\rangle \rmd t +\frac
  1{2\epsilon}\sum_{k=1}^{N}D^{2}(|\psi(t_{k})\rangle, |\psi_{k}\rangle)
\end{equation}
where  $|\psi(t)\rangle=U(t)|\psi_{0}\rangle$ represents
the trajectory in ${\mathcal H}$ defined by the unitary evolution $U(t)$,
the scalar product $\langle \cdot \mid   \cdot\rangle$ represents the
trace of the product 
    \begin{equation}
      \label{eq:1}
      \langle A\mid B\rangle=-\frac12\mathrm{Tr}(AB), \qquad \forall
      A, B\in \mathfrak{u}(\mathcal{H}), 
    \end{equation}
and $\epsilon$ is just a positive coefficient which allows us to change
the relative importance of the total change of the Hamiltonian of the
system and how close we are from the point we are supposed to target
($D$ represents the corresponding distance). 

\smallskip

Some comments are in order:
\begin{itemize}
  \item As the time dependence of the Hamiltonian in quantum control
    problems is often implemented as time-dependent magnetic fields
    which interact with the spin system, we can think at the
    minimization problem as the equilibrium between the accuracy at
    targetting the points at the designated times and the total energy
    required to change the magnetic fields during the process.
\item The (square) distances $D^{2}(|\psi(t_{k})\rangle, |\psi_{k}\rangle)$
are considered on the projective space ${\mathcal PH}$, with respect to
the canonical Fubini-Study metric. From the technical point of view,
the functional (\ref{eq:JproblemA}) combines thus quantities defined
on two different manifolds: the trace-norm of skew-Hermitian operators and
the Fubini-Study metric on the projective space.
\item The solution of the problem above, even for very simple examples
  such as  a
  two-level quantum system, requires of numerical algorithms. In
  \cite{Brody2012} the authors provide an algorithm for the
  discretization of their variational equations which is used in the
  construction of the solution of the example they study.
\end{itemize}

The aim of this paper is to introduce an alternative formulation of
the problem of quantum splines and a generalization of the notion
which enlarges the range of potential applications. The main idea is to
re-formulate the problem in terms of density matrices and to adapt the
conditions on the dynamics to it. Thus, the notion of quantum spline
for pure states introduced in \cite{Brody2012} is  included, while
an analogous notion for mixed states does also make sense:

\vspace*{0,5cm}

\fbox{
\begin{minipage}{0.9\textwidth}
\textbf{New formulation of the interpolation problem}: Let us consider a set of density matrices
$\{ \rho_{0}, \cdots  ,\rho_{N}\}$ on a certain finite dimensional
Hilbert space ${\mathcal H}$ and a corresponding set of times $\{t_{0},
\cdots, t_{N}\}$. 
Find a time-dependent Hamiltonian $H(t)$ defining an unitary evolution
which, at the given times $t_{k}$,  passes as close as
  possible  to the
given points $\rho_{k}$, and such that the change in the Hamiltonian
is optimal in the sense that the cost functional
\begin{equation}  \label{eq:JproblemB}
  {\mathcal J}=\sum_{{k=1}}^{N}\int_{t_{k-1}}^{t_{k}}\langle \rmi\dot H(t) \mid \rmi\dot
  H(t)\rangle \rmd t +\frac
  1{2\epsilon}\sum_{k=1}^{N}d^{2}(\rho(t_k), \rho_{k})
\end{equation}
is minimized, where the trajectory of the system  represented by
$\rho(t)$ is a solution of von Neumann equation: 
\begin{equation} \label{eq:vonNeumann}
\rmi\hbar\dot \rho(t)=H(t)\rho(t)-\rho(t)H(t),
\end{equation}
and $d$ represents the distance on the space of Hermitian operators

$d(A,B)=\sqrt{\frac 12\mathrm{Tr}(A-B)^{2}} $.
\end{minipage}
}

\vspace*{0,5cm}

This formulation of the problem includes the previous one for the
case of pure quantum states, but it allows to consider it
also for general quantum states.  Furthermore, it relaxes the continuity
condition on the derivative of the Hamiltonian $H(t)$ at the
intermediate points, as it is often assumed in (quantum)
control based on pulses. From a physical point of view, the lack of continuity does
not represent a serious condition on the system, and as we are
going to see it allows for a much simpler analysis of the problem.

Regarding the generalization to density matrices, notice that all real
systems are in contact with some 
  type of environment. In that case, we know that the system is very
  seldom in a pure quantum state and therefore to extend the notion of
  quantum spline to mixed states seems as a natural step forward.
For instance,  we could consider a system which
initially is   in equilibrium at a finite temperature $T$ (see \cite{Breuer2002}) with a
bath. If the evolution associated to the quantum controls is much
faster than the dynamics produced by the interaction with
the bath, we could approximate the evolution by a unitary
transformation on the set of mixed quantum states.  This is the type
of problem that we will be considering in this paper.  Another
interesting generalization would be to consider that the controlled
evolution is of the same order as the interaction with the bath and
replace von Neumann equation by a more general master equation, such
as Lindblad-Kossakowski equation. The problem would
be formally analogous to our ``New formulation'' above, replacing equation
(\ref{eq:vonNeumann}) by Lindblad-Kossakowski equation.  Nonetheless,
Lindblad-Kossakowski equation would put some 
constraints in the possible values of the set of times $\{t_{j}\}$ and
density matrices $\{\rho_{j}\}$ (remember that  Lindblad dynamics is
not always controllable \cite{Altafini2002, Altafini2003}).  We are
also studying this problem and it will be considered in a future paper. 

There are important aspects that characterize the new formulation:

\begin{itemize}

\item
 From the technical point of view, our formulation is much simpler
 than the one considered in \cite{Brody2012}, since it can be done at
 the level of a linear space (the dual space to the algebra of the
 unitary group $\u^{*}(n)$), instead of considering the Lie group $U(n)$ and the
 projective space $\mathbb{CP}^{n-1}$. 

 \item The set of quantum states $\mathcal{D}$ for our formulation
   will be the submanifold of  $\u^{*}(n)$ defined by the elements which
   have trace equal to one and are positive definite. $\mathcal{D}$ is
   a nonlinear submanifold of $\u^{*}(n)$, but the particular choice
   of our dynamic system allows us to treat the problem as a problem
   defined on the linear space $\u^{*}(n)$. Indeed, as the dynamics
   on $\mathcal{D}$ is defined by von Neumann
   equation, we know that the  solution $\rho(t)$ must belong to a
   unitary orbit of the coadjoint action of the unitary group. As it
   is well known (see \cite{Konstant1970}), those  orbits constitute symplectic submanifolds of
   $\u^{*}(n)$ which define the leaves of the symplectic foliation
   associated with its Lie-Poisson canonical structure. If we consider
   an initial condition contained in one of the leaves, the whole
   solution will be contained in it.  This property will allow us to
   consider the problem defined directly at the linear space
   $\u^{*}(n)$ and forget about the nonlinear constraints.
   Nonetheless, this also imposes some constraints for
     the definition of splines in the general case. Indeed, as the
     dynamics we are considering is always unitary, any generalized
     spline will be contained in a unitary orbit. Thus, if the target points
    are not contained in one unitary orbit, the best possible solution
  will only be able to define the closest unitary orbit to the set of
  target points. We will discuss this point at the end of the paper.
 \item The geometric
   formulation of Quantum Mechanics (see \cite{Ashtekar:1997p906,Brody2001,Carinena2007b,Kibble:1979p7279} ) is  a
   natural framework to formulate the problem in a geometric formalism
   which exhibits these symplectic aspects. In that framework, von Neumann equation
   defines in a natural way a Hamiltonian vector field. This
   property is particularly useful when considering numeric
   integration of the solutions, since by using symplectic integrators
   we do not need to take into account constraints such as the trace
   of the quantum state or its rank and the problem can be treated as
   a free one.   
\item In  \cite{Brody2012}, a numerical algorithm was introduced in
  order to provide a method of integration for the splines. In our
  formalism, even if simpler, a numerical algorithm will also be
  necessary. We introduce an iterative algorithm to
  approach the optimal solution of the
  interpolation problem. We may not reach the absolute optimal solution, but we
  always obtain a very good approximation to it. As our framework is
  defined on linear spaces   the algorithm turns out to be  very
  simple and the   efficiency is   very high.   
\end{itemize}

The structure of the paper is as follows. Section \ref{Section2}
presents a summary of the geometric formalism of quantum mechanics,
with special emphasis on the geometrical structures associated with
the description of the 
set of Hermitian operators and the set of density states, which are
the main ingredients of our construction. Section \ref{Section3} presents
the main aspects of our new formulation, analyzing first the
quantum control problem 
from the point of view of the Pontryagin maximum principle, then the
analytical tools used to build the system of differential equations
defining the solution and finally discussing an iterative algorithm to
build solutions in an efficient way and the numerical methods 
we use to obtain them. Section \ref{Section4} illustrates our
construction with the simplest example for the interpolation between a
set of pure states of a two level system and Section \ref{Section5}
presents a more sophisticated example of the interpolation of mixed
states  of a three level quantum systems. Finally,  Section \ref{Conclusions} summarizes the main
results of the paper and  discuss the possible
generalization of our method to the case of open quantum systems.

\section{The geometrical description of quantum mechanics}
\label{Section2}

{\color{black}

Let us  briefly review  some very general aspects of the geometrical formulation of quantum mechanics, focusing on the ingredients which will be used later in the paper. For more details, we address the interested reader to  \cite{Ashtekar:1997p906,Brody2001,Carinena2007b,Kibble:1979p7279} and references therein.  We use Einstein summation over repeated indexes.


\subsection{The geometrical structures of $\mathfrak{u}^{*}(n)$ }

We hereafter assume $\Hil$ to be an $n$-dimensional complex Hilbert space.
In this case the $C^*$--algebra of operators corresponds to
$\mathcal{A}=\mathrm{End}(\C^{n})$ and the involution is the usual
adjoint operation for complex endomorphisms.  In this case, when
considering an orthonormal basis, self-adjoint operators are
represented by Hermitian matrices. 

\bigskip

It is important now to emphasize  the following facts:
\begin{itemize}

\item 
We can define  isomorphisms between $\u(n)$, $\u^*(n)$ and the vector space of
Hermitian matrices $\rmi \u(n)$ by  mappings
\begin{equation}
  \label{eq:3}
  \rmi \u(n) \ni A \mapsto iA\, \in \u(n);
\end{equation}
\begin{equation}
  \label{eq:2}
\rmi \u(n) \ni A \mapsto \xi_A\, \in \u^*(n),\;\;\;\mbox{with}\;\; \xi_A=\langle -\rmi A \mid \cdot\, \rangle: \u(n)\to \mathbb{R}.
\end{equation}
These isomorphisms define also Lie brackets on each space $[\cdot,
\cdot]_{\u}$, $[\cdot, \cdot]_{\u^{*}}$ and $[\cdot,
\cdot]_{\rmi \u}$.
We have a special interest in the identification of  the
elements of  $\u^*(n)$ with the Hermitian matrices, which will be used later.
\item  On the dual space $\u^*(n)$ we have a natural linear Poisson
  structure, the Lie-Poisson structure.  The  Poisson tensor $\Lambda$
  is defined,  on the set of linear functions on $\u^*(n)$, as follows
  \begin{equation}
    \label{eq:4}
\Lambda_{\xi}(V,W)=\xi \left([ V, W]\right), \qquad
\xi\in \u^*(n), \quad \forall  V, W\in  \u(n).
\end{equation}
From this definition on linear functions, we can construct
the  Poisson bracket on general functions  $f, g \in {\mathcal
  C}^{\infty}(\u^*(n))$ by extending  equation (\ref{eq:4}) requiring
bilinearity, i.e.,  as the action of the  bi-differential operator
which on linear functions behaves as equation (\ref{eq:4}). This
defines the canonical Lie-Poisson tensor of the dual of the unitary
algebra (see \cite{Abraham1978}). 
\item 
The Hamiltonian vector field  $X_{\rmi H}=-\Lambda(\rmi H, \cdot) $
is the infinitesimal generator of the coadjoint action of $U(n)$ on
$\u^*(n)$ and, therefore,  the  system
\begin{equation}
\label{eq:LP}
\dot\xi(t)= X_{\rmi H}(\xi(t)), \qquad \xi(0)=\xi_0, \qquad
\xi(t)\in \u^*(n),
\end{equation} 
is a symplectic Hamiltonian system on the coadjoint  orbit where the
initial condition $\xi_0$ is contained (again we refer the reader to
\cite{Abraham1978} for details) .

\end{itemize}

\bigskip

\subsection{The set of density matrices ${\mathcal D}({\mathcal H})$}
The physical states which are used in the Schr\"{o}dinger formalism
correspond  to the points of the projective space
${\mathcal PH}$. As it is well known, we can  also represent these states by using 
the projectors on one-dimensional subspaces of the Hilbert
space. Unitary evolution, associated with the
Schr\"{o}dinger or the von Neumann equation,  will define trajectories on this set. 
 We will denote  by ${\mathcal  D}^{1}({\mathcal H})$ the corresponding set of
 projectors. Nonetheless,  ${\mathcal  D}^{1}({\mathcal H})$ is not enough to
represent all the possible physical states of a system, since
arbitrary convex combinations of rank-one-projectors also define
admissible physical states.  Indeed, if our system is not isolated and
is surrounded by an environment (as it is the case of all real
systems), or just a  set of other systems, the representing state will
not be, in general, pure. Therefore, we  must enlarge the set of
states to consider: 

\begin{definition}
The set of density states ${\mathcal D}({\mathcal H})$  of the system
corresponds to the subset of $\u^*(n)$ obtained by convex
combinations of rank-one-projectors, that is,
\begin{equation}
  {\mathcal D}({\mathcal H})=\left \{ \rho= \sum_{k} p_k \rho_k \; :\quad p_k\geq 0, \quad  \sum_{k}
  p_k=1, \quad \rho_k\in {\mathcal D}^1({\mathcal H})\right \}.
\end{equation}
Equivalently, we can consider the following definition: an element
$\rho\in  \u^*(n)$ is a density operator if and only if 
\begin{equation}
  \Tr \rho=1 \quad \mbox{and}\quad\rho\geq 0.
\end{equation}
\end{definition}


An important property of the manifold $\mathcal{D}(\mathcal{H})$
is its internal structure, in particular in what regards the
restriction of the coadjoint orbits of the unitary group on
$\u^{*}(n)$. A very important aspect of that structure is the nature
of $\mathcal{D}(\mathcal{H})$ as a stratified manifold, the strata
being defined by the rank of the states (see \cite{Grabowski2005a} for
details).  Indeed, as all density states are self-adjoint and positive
definite, they all
are diagonalizable and have non-negative eigenvalues.  The number of
non-vanishing eigenvalues equals the rank of the operator (or the
matrix, if we choose a basis).  Thus, all states having the same rank
belong to the same stratum. As the coadjoint action of the unitary
group is known to preserve the spectrum of the density operator, it is
obvious that the orbit will stay on the same stratum as its initial
condition. Equivalently, we can claim that the Hamiltonian vector
field defined by equation (\ref{eq:LP}) will be tangent to the
different strata.

Thus, if we consider the Hamiltonian vector field $X_{\rmi H}$
(defined by equation (\ref{eq:LP}))
restricted to $\mathcal{D}(\mathcal{H})$, we obtain the von Neumann
equation for the density matrix  $\rho\in \mathcal{D}(\mathcal{H})$, 
\begin{equation} \label{HamVecField}
  \dot \rho=X_{\rmi H}(\rho).
\end{equation}
  We know that this evolution corresponds to a unitary
  transformation. Thus, the Hamiltonian vector fields of $\Lambda$ are
  the infinitesimal generators of  the coadjoint action of the unitary
  group on $\u^*(n)$, and the 
corresponding orbits are known to be symplectic submanifolds.  For
instance,  the set of pure states $\mathcal{D}^{1}(\mathcal{H})$
defines a symplectic orbit which corresponds to a symplectic leaf of
the canonical Poisson 
foliation. Therefore, the solution of von Neumann's equation on
$\mathcal{D}^{1}(\mathcal{H})$ is always tangent to it and defines  a
symplectic transformation. This allows us to use the formulation of
the problem on  $\mathfrak{u}^*(n)$ without using any constraints to specify the
submanifold of pure states, since we know that any trajectory with an
initial condition on $\mathcal{D}^{1}(\mathcal{H})$ will remain on the
submanifold for  all times.

For the case of a mixed state, the situation is analogous, but in this
case the symplectic orbit of the coadjoint action will not cover all
the stratum.  Each stratum will contain several coadjoint orbits of
different dimensions  and
therefore the formulation of the interpolation problem must take into
account this property. Indeed, the exact quantum spline for mixed
states (i.e., a curve joining all points) makes sense only  if the set
of points belong to the same coadjoint orbit.  As the target points
are not reached exactly, the problem still makes sense for a
neighborhood of the orbit. In the next Section we will see how it is
possible to define an iterative algorithm which drives the evolution
to the closest point of the orbit with respect to the target point.

}

\section{ Dynamical interpolation problem on $\u^*(n)$}
\label{Section3}

Let $\{ \rho_0, \rho_{1}, \cdots, \rho_{N}\}$ be a finite subset of
$\D(\mathcal{H})$ and $0=t_0<t_1<\cdots<t_N=T$ a partition of the time
interval  $[0,T]$, $T>0$. Consider also a $m$-dimensional vector
subspace  $\h$ of $\u^*(n)$, $m\leq n^2$ {\color{black} which will
  represent the space of admissible controls for our interpolation problem}. 

\smallskip


\subsection{The control problems chain }

In general, let $E$ be a real vector space of dimension $n$ and $U$ a
subspace of $E$ (in our case, $E$ will be the vector space
 $\u^*(n)\times \u^*(n)$). Given a cost function $L:E\times U\to \mathbb{R}$  and the
functions $G_j:E\to \mathbb{R}$, consider the optimal control  problem
P concerning the minimum of the functional 
\begin{equation}
\displaystyle J(u)=\sum_{j=1}^{N}\int_{t_{j-1}}^{t_{j}}L(x(t),u(t))\,
\rmd t+\frac 1{\epsilon}\sum_{j=1}^N G_j(x(t_j)), 
\end{equation}
 where   $\epsilon$ is a tunable positive parameter, and  with
 $u:[0,T]\to U$  and $x:[0,T]\to E$  being  curves, 
subject to the initial condition
$x(0)=x_0,$
satisfying  the  control system
\begin{equation}
\dot x =f(x,u)
\end{equation}
and the conditions:

\begin{itemize}
	
	\item[--]  $x$ is continuous in $[0,T]$ and smooth in
          $]t_{j-1},t_j[$, with finite {\color{black} one-sided}  limits at $t_j$, for
          all $j=1, \ldots, N$; 
	
	\item[--] $u$ is smooth in $]t_{j-1},t_j[$, with finite 
           {\color{black} one-sided}  limits at $t_j$, for all $j=1, \ldots, N$.   
\end{itemize}
	
	This defines a set of $N$ subproblems $P_j$, $j=1, \ldots, N$,
        one for each subinterval, where the initial condition is fixed
        (by the value 
of the solution at the end point of the previous interval), but the
final condition is free. {\color{black}Thus, on interval $]t_{j-1},t_j[$ we consider
the functional 
\begin{equation}
  \label{eq:11}
\displaystyle J_{j}(u)=\int_{t_{j-1}}^{t_{j}}L(x(t),u(t))\,
\rmd t+\frac 1{\epsilon} G_j(x(t_j), 
\end{equation}
and search for the optimal solution $x:[t_{j-1},t_{j}]\to \R$ subject
to the initial conditions $x(t_{j-1})=x_{j-1}$,
which corresponds to the final conditions of the previous subinterval.  }
Each of these subproblems is a particular
case of a well known optimal control problem known as a Bolza
problem (see \cite{Cesari1983}).
The solution $x(t)$   is asked to be continuous on the whole interval 
$[0,T]$.   

The solution of each  Bolza problem can be obtained 
straightforwardly by using   the Pontryagin maximum principle, the
only difficulty being the boundary condition. Although it is a well
known method, let us present, for completeness,  a sketch of the
variational construction needed to prove Pontryagin's result. Let us 
take, for instance,  the first subinterval $[t_0, t_1]$.      
Consider the Hamiltonian function $h: E\times E^* \times U
\to\mathbb{R}$  associated with the optimal control problem, defined
by     
\begin{eqnarray}
h(x,p,u)=p(f(x,u))-L(x,u).
\end{eqnarray}

For each $x\in E$ and $p\in E^*$, the partial functional derivative of
a function $h$ with respect to $x$ at $a$  is an element of $E^*$
denoted by $(\delta h/\delta x)(a)$ and  the partial functional 
derivative with  
respect to $p$ at $a$ is an element of $E$  denoted by $(\delta h/\delta p)(a)$.
Now, let us enlarge the  functional $J$  by the dynamical constraints,
using the costate trajectory $p:[t_0,t_1]\to E$, in the following way: 
\begin{eqnarray*}
\widetilde{J}(u)
&=& \displaystyle \int_{t_{0}}^{t_{1}}L(x(t), u(t))\,\rmd t
+\int_{t_{0}}^{t_{1}} p\left(\dot x-f(x,u)\right)\,\rmd t + \frac1{\epsilon} G_1(x(t_1))
\nonumber \\[15pt]
&=&  \displaystyle \int_{t_{0}}^{t_{1}}\left(L(x,
    u)-p(f(x,u))\right)\rmd t+\int_{t_{0}}^{t_{1}} p(\dot x)\,\rmd
    t+\frac1{\epsilon} G_1(x(t_1)) 
\\[15pt] 
&=& \displaystyle  -\int_{t_{0}}^{t_{1}} h(x,p,u)\rmd t+\int_{t_{0}}^{t_{1}} p(\dot x)\,\rmd t+\frac1{\epsilon}G_1(x(t_1)). 
\end{eqnarray*}
Thus, we know that the optimal solution will satisfy the equations 
\begin{equation}
\hspace{-1.5cm}\dot x=\frac{\delta h}{\delta p},\qquad \dot p=\displaystyle-\frac{\delta h}{\delta x}, \qquad
p(t_1^-)=-\frac1{\epsilon}\frac{\delta G_1}{\delta x}(x(t_1)), \qquad
\frac{\delta h}{\delta u}(x,p, u)=0. 
\end{equation}

\bigskip

Therefore, it is now simple to prove the following result:

\begin{theorem}
  \label{teor:General}
	
If $u$ is the optimal control resulting of the optimal control problems $P_j$,   $j=1, \ldots, N$
	and $x$ is the associated optimal state trajectory, then there exists a piecewise-smooth costate trajectory $p: [0,T]: \rightarrow E$ such that
	\begin{eqnarray} 
	\displaystyle\dot x=\frac{\delta h}{\delta p},  \qquad   \dot p=-\frac{\delta h}{\delta x}, \label{eq:GenHamSystem0}
	\\[10pt] \frac{\delta h}{\delta u}=0,  \label{eq:GenHamSystemU0} \\[5pt]
	\displaystyle p(t_j^-)=-\frac1{\epsilon}\frac{\delta G_j}{\delta x}(x(t_j)),\qquad\mbox{for all}\quad j=1, \ldots, N.
	\end{eqnarray} 
\end{theorem} 

\smallskip

	{ \bf Proof}
	
The problem is defined piecewisely, and hence we will proceed in the
same way.  For each $j$, consider the problem on the subinterval
$[t_{j-1}, t_j]$, determine the  optimal solution on it, and consider
then the value of the 
optimal solution to define the boundary condition of the next
subinterval in order to ensure continuity of the trajectory.  On the
first subinterval, $[t_0, t_1]$, we have seen that the  solution
satisfies the condition. Consider now the optimal solution $x(t)$ and
the corresponding value $x(t_1)$, and use it as the initial condition
for the Bolza problem in the subinterval  $[t_1, t_2]$. We will then
define an optimal solution for the second subinterval which 
will define the initial condition for the third one. On each case we produce a smooth costate variable $p$, defined on each subinterval.   If we repeat the procedure on all subintervals, the result follows and the global
costate variable will be, by construction,  piecewise-smooth. \quad$\Box$

\bigskip

Observe that system  (\ref{eq:GenHamSystem0}-\ref{eq:GenHamSystemU0})  and the time-invariance of the Hamiltonian $h$ guarantee that $h$ is an integral of motion.

\subsection{The optimal control problem P for quantum splines}

Let us apply now the result above to our formulation of the
interpolation problem. As we had formulated it on $\u^*(n)$,
we consider a cost function  \mbox{$L:\u^*(n)\times \u^*(n)\times \mathfrak{h}\to \mathbb{R}$} defined by
\begin{equation} \label{eq:CostFunction}
L(\rho,H,u)=\frac{1}{2}\norm u ^2=\frac 12 \langle u, u\rangle
\end{equation}
{\color{black} where $\langle \xi, \eta \rangle=\frac 12 \mathrm{Tr}(\xi \eta)$ for $\xi, \eta\in
\u^{*}(n)$. The functions $G_j:\u^*(n)\to \mathbb{R}$
are given by}
\begin{equation}
G_j(\rho)= \frac 1{2}\norm{\rho-\rho_j}^2, \qquad j=1, \ldots, N.
\end{equation}

The dynamical interpolation problem is the optimal control problem
concerning the minimum of the functional
\begin{equation}
  \label{eq:9}
 J(u)=\sum_{j=1}^{N}\int_{t_{j-1}}^{t_{j}}L(\rho(t),H(t),u(t))\, \rmd t+\frac 1{\epsilon}\sum_{j=1}^N G_j(\rho(t_j)),
\end{equation}
with $u:[0,T]\to \mathfrak{h}$  and $(\rho,H):[0,T]\to \u^{*}(n)\times
\u^*(n)$  being piecewise-defined curves subject to the initial
conditions 
\begin{equation}\label{eq:InitialC}
\rho(0)=\rho_0 \qquad  \mbox{and} \qquad H(0)=H_0,
\end{equation}
satisfying  the  control system
\begin{equation}
  \dot \rho ={\color{black}[H,\rho]_{\u^{*}}}, \qquad
     \dot H=u,
\end{equation}
where the curves verify the same regularity conditions of the previous
section. 

The Hamiltonian function to apply the Pontryagin maximum principle
becomes now the function $h: \u^*(n)\times \u^*(n)\times \u(n)\times
\u(n)\times \mathfrak{h}\to \mathbb{R}$ given by 
\begin{equation} 
h(\rho,H,\Gamma,\Pi,u)=\Gamma({\color{black}[H,\rho]_{\u^{*}}})+\Pi(u)-L(\rho,H,u). 
\end{equation}

\smallskip 

The approach considered above, obtained  by application of the theorem
\ref{teor:General}, implies the following result:

\begin{theorem}
If $u$ is the optimal control resulting of the optimal control
problems $P_j$, $j=1, \ldots, N$,  and $(\rho,H)$ is the associated
optimal state trajectory, then there exists a piecewise-continuous
optimal costate trajectory  $(\Gamma, \Pi)$  that 
satisfies the  system
\begin{equation}\label{eq:Hamilton}
   \dot \rho={\color{black}[H,\rho]_{\u^{*}}}, \quad
  \dot H=u, \quad
  \dot{\Gamma}=-{\rm ad}_{H}^*\Gamma, \quad
  \dot\Pi=\,{\rm ad}_{\rho}^*\Gamma, \quad
 \Pi=u^{\flat},
\end{equation}
the initial conditions (\ref{eq:InitialC}) and the interpolating conditions
\begin{equation}
\Gamma(t_j^-)=-\frac{1}{\epsilon}(\rho(t_j)-\rho_j)^{\flat},\qquad
  \Pi(t_j^-)=0, \qquad \,j=1,\ldots,N.  
\end{equation}
{\color{black} In the equations above $\mathrm{ad}$ and $\mathrm{ad}^{*}$ refer to
the adjoint and co-adjoint actions of the Lie bracket $[\cdot,
\cdot]_{\u^{*}}$ and ${\cdot}^{\flat}$ specifies the isomorphism
$\flat:\u^{*}\to \u$ given by the inner product on $\u^{* }(n)$}.
\end{theorem}


Notice that eliminating the controls in (\ref{eq:Hamilton}) we obtain the system
\begin{equation}\label{eq:HSywithouU}
\dot \rho={\color{black}[H,\rho]_{\u^{*}}}, 
   \qquad \dot H=\Pi_{\h}^{\sharp} ,\qquad
\dot{\Gamma}=-{\rm ad}_{H}^*\Gamma, 
    \qquad  \dot\Pi=\,{\rm ad}_{\rho}^*\Gamma,
\end{equation} 
 which is Hamiltonian  with respect to the function $\tilde
 h:\u^*(n)\times\u^*(n)\times\u(n)\times\u(n)\to \mathbb{R}$, 
\begin{equation}
\tilde h(\rho,H,\Gamma,\Pi)
=\Gamma({\color{black}[H,\rho]_{\u^{*}}})+\frac{1}{2}\norm{\Pi_{\h}^{\sharp}}^2.  
\end{equation}
{\color{black}Notice that ${\cdot }^{\sharp}$ specifies the inverse of the  isomorphism $\flat$.}
\smallskip

Observe that the Hamiltonian system (\ref{eq:HSywithouU}) can be written as follows
\begin{equation} 
\label{eq:HSwithoutUwithAD}
    \dot \rho={\color{black}[H,\rho]_{\u^{*}}},
   \qquad \dot H=\Pi_{\h}^{\sharp}, \qquad 
   \dot{\Gamma}={\color{black}[H,\Gamma^{\sharp}]_{\u^{*}}^{\flat}},
    \qquad  \dot\Pi={\color{black}[\Gamma^{\sharp},\rho]_{\u^{*}}^{\flat}},
\end{equation}
with interpolation conditions
\begin{equation}
\Gamma(t_j^-)=-\frac{1}{\epsilon}(\rho(t_j)-\rho_j)^{\flat}, \qquad \Pi(t_j^-)=0, \qquad j=1, \ldots, N.
\end{equation}

\bigskip

Consider now the problem $P$ with $u$ taking values in all space, that
is, {\color{black} the space of admissible controls corresponds to
  $\h=\u^{*}(n)$ and thus $m=n^{2}$}. From Hamiltonian 
equations (\ref{eq:HSwithoutUwithAD}), 
we differentiate  twice the equation $\dot H=\Pi^{\sharp}$ and
eliminate the 
costates $\Gamma$ and $\Pi$, to obtain the equations  $\ddot
H=-\mathrm{ad}_{\rho}\Gamma^{\sharp}$ and $\dddot{H}={\color{black}[H,\ddot H]_{\u^{*}}}$. Indeed,
\begin{equation}
\begin{array}{lll}
\displaystyle  \dddot{H}&=&-\,{\rm ad}_{\dot \rho}\Gamma^{\sharp}-\,{\rm ad}_{\rho}\dot \Gamma^{\sharp}
= -{\rm ad}_{[H,\rho]_{\u^{*}}}\Gamma^{\sharp}-\,{\rm ad}_{\rho}{\rm ad}_{H}\Gamma^{\sharp}\\[5pt]
&=& {\rm ad}_{\rho}{\rm ad}_{H}\Gamma^{\sharp}-{\rm ad}_{H}{\rm ad}_{\rho}\Gamma^{\sharp}-\,{\rm ad}_{\rho}{\rm ad}_{H}\Gamma^{\sharp}
\\[5pt]
&=&  -{\rm ad}_{H}{\rm ad}_{\rho}\Gamma^{\sharp}= {\rm ad}_{H}\ddot H={\color{black}[H,\ddot H]_{\u^{*}}}.
\end{array}
\end{equation}
Therefore,  the associated optimal state trajectory $(\rho,H)$ verifies the two conditions
\begin{equation}\label{eq:DCP}
\dot \rho={\color{black}[H,\rho]_{\u^{*}}} \qquad \mbox{and}\qquad \displaystyle
\dddot{H}={\color{black}[H, \ddot H]_{\u^{*}}}.
\end{equation}
Moreover, interpolation conditions can be interpreted in terms of $\rho$ and $H$ as follows: $\rho$ is of class $\mathcal{C}^1$, $H$ is of class $\mathcal{C}^0$ and
\begin{equation}
  \label{eq:8}
  \ddot{H}(t_j^-)=\frac 1{\epsilon}{\color{black}[\rho_{j},\rho(t_j)]_{\u^{*}}},\qquad \dot{H}(t_j^-)=0, \qquad \mbox{for all}\quad j=1, \ldots, N.
\end{equation}

\smallskip

Now, if we integrate the equation $\displaystyle \dddot{H}=[H,\ddot H]_{\u^{*}}$, we obtain
\begin{equation}
  \label{eq:7}
  \ddot H(t)=K_j+{\color{black}[H(t), \dot H(t)]_{\u^{*}}},\qquad\mbox{with}\quad t\in [t_{j-1},t_j],
\end{equation}
where $K_j$ are the matrices defined by
\begin{eqnarray}
\label{eq:6}
  K_j=\frac 1{\epsilon}{\color{black}[\rho_{j},\rho(t_j)]_{\u^{*}}}, \qquad \mbox{for all}  \quad j=1,\ldots,N.
\end{eqnarray}

{\color{black}Notice that any solution of equations (\ref{eq:7}-\ref{eq:6}) satisfies the second
equation of (\ref{eq:DCP}), i.e., any of them are cubic polynomials
and represent extremals of  the continuous part of functional $J$
defined in equation  (\ref{eq:9}). 
}

Let $\{\sigma_{l}\mid l=1,\ldots,n^2\}$ be  an orthonormal basis of
$\u^*(n)$. Denote, with respect to this basis,  the coordinates  of
$K_j$ by $(k_j^l)$, $l=1,\ldots,n^2$, and the coordinate functions for
$\rho$ and $H$ 
by $(x_l)$ and $(y_l)$, $l=1,\ldots,n^2$, respectively. The system
(\ref{eq:DCP}) is {\color{black} then}  written in the following way 
\begin{equation}
  \label{eq:coordinates}
\begin{array}{l}
\displaystyle  \dot x_l(t)=\sum_{r,s=1}^{n^2}c^l_{rs}\,y_r(t) x_s(t), \\[15pt]
\displaystyle  \ddot y_l(t)=k_j^l+\sum_{r,s=1}^{n^2}c^l_{rs} \,y_r(t) \dot y_s(t),
\end{array}
\end{equation}
for $l=1,\ldots,n^2$ and $t\in [t_{j-1},t_j]$, with $j=1,\ldots,N$ and
where $c^l_{rs}$ are the  structure constants of the Lie algebra
{\color{black} structure of } $\u^*(n)$.

{\color{black}
\subsection{Implicit numeric integration}
\label{sec:impl-numer-integr}
The integration of equations (\ref{eq:coordinates}) is  far from trivial  because of
the presence of the coefficients $\{k_{j}^l\}$, which require of the
solution of $\rho(t)$ to be determined, since
$$
K_{j}=\frac 1\epsilon[\rho_{j}, \rho(t_{j})]_{\u^{*}}.
$$
These operators couple the equation for $\rho(t)$ and the equation for
$H(t)$, which, otherwise, would be independent.  In any case, it is
not clear whether the optimal solution exist. Therefore, 
we have designed an iterative  algorithm to obtain a numerical
solution for equations  (\ref{eq:DCP}), always improving the distance
to the target points. We consider only one  interval $[t_{j-1},
t_{j}]$, for the sake of simplicity:
\begin{itemize}
\item We fix the initial value of $K_{j}=K^{0}_{j}=0$ (the
  null matrix) and integrate equation
  (\ref{eq:coordinates}). This gives us an initial solution $\rho^{0}(t)$ which, in
  general, will be a  bad solution of the interpolation problem
  from $\rho_{j-1}$ to $\rho_{j}$.
\item From this solution we determine the corresponding value of
  $K^{1}_{j}:=\frac 1\epsilon[\rho_{j}, \rho^{0}(t_{j})]_{\u^{*}}$. Notice that this
  expression corresponds to the extremal of the variations of 
  the cost functional  $G=\frac 1{2\epsilon} \| \rho-\rho_{j}\|^{2}$, and
  therefore it defines the suitable steering acceleration of the
  solution $(\rho(t),H(t))$ to make $\rho(t_{j})$ to move towards
  $\rho_{j}$.  For a suitable (not too small) value of $\epsilon$, the
  computation of  a new solution of equation  (\ref{eq:coordinates}) with this value of $K_{j}^{1}$,
  produces  a  solution $(\rho^{1}(t), H^{1}(t))$, where $\rho^{1}(t)$
  is a  better solution of the    interpolation problem. If the value
  of $\epsilon$ is too small, the 
  acceleration of $H^{1}(t)$, even in the correct direction,  will be too
  large and the solution $\rho^{1}(t)$ might  become a worst
  interpolation solution than $\rho^{0}(t)$. 
 
\item From solution $\rho^{1}(t)$  we recompute  
  $K^{2}_{j}:=K^{1}_{j}+\frac 1\epsilon[\rho_{j}, \rho^{1}(t_{j})]_{\u^{*}}$ and integrate
  again  equation   (\ref{eq:coordinates}) to define a new solution
  $(\rho^{2}(t), H^{2}(t))$, where $\rho^{2}(t)$  will be a better
  solution of the interpolation 
  problem.   We repeat the process to obtain a sequence of solutions
  $\{\rho^{j}(t)\}$, each one representing a better interpolation
  solution for $\rho(t)$.  Notice that any curve of this series is a
  solution of equation  (\ref{eq:DCP}) and therefore provides a
  solution for the interpolation problem.  Each iteration
  allows us to obtain solutions of equations  (\ref{eq:DCP}) which are
  closer and closer to the target points.
At the same time, the number of iterations
  of the algorithm also affects the value of the continuous part of
  the functional $J$ (equation \ref{eq:9}): the larger the number of
  iterations, the larger value for the integral part. Hence adjusting  the number
  of iterations  can also be considered a mechanism
  to obtain solutions of the interpolation problem which assign
  different  weights  to the  continuous part with respect to the
  discrete part of   the functional. 
\item On each of these steps the value of
  $\frac 1\epsilon[\rho_{j}, \rho^{i}(t_{j})]_{{\u^{*}}}$ 
  decreases, since $\| \rho_{j}-\rho^{i}(t_{j})\|^{2}$ is
  decreasing. Therefore, after a sufficiently large number $r$ of
  steps, the value of $K_{j}^{r}$ stabilizes within a certain
  tolerance.  The speed of the process   depends on the value of
  $\epsilon$, that is fixing the magnitude of 
  $K_{j}$  (i.e.  an acceleration for $H(t)$) and thus fixes
  the velocity of von Neumann equation for $\rho(t)$.  In any case,
  notice that $(\rho^{i}(t), H^{i}(t))$ is  not  a solution of equation
  (\ref{eq:6}) since $H^{i}(t)$ is a solution of an
  equation obtained with respect to a value of $K^{i}_{j}$ which has been
  obtained from $\rho^{i-1}(t)$ as

  \begin{equation}
    \label{eq:10}
  K^{i}_{j}=K^{i-1}_{j}+\frac 1\epsilon[\rho_{j},
  \rho^{i-1}(t_{j})]_{{\u^{*}}}=\frac 1\epsilon \left [\rho_{j},
    \sum_{s=0}^{i-1}\rho^{s}(t_{j}) \right ]_{{\u^{*}}},
     \end{equation}
   and not as
  $$
K_{j}=\frac 1\epsilon[\rho_{j}, \rho^{i}(t_{j})]_{\u^{*}},
$$
which is the form required for the solution $(\rho^{i}(t), H^{i}(t))$ to
be a solution of equations (\ref{eq:coordinates}) and
(\ref{eq:6}). Hence we can conclude that it is not the optimal
solution for our problem. Nonetheless,  it is still  a solution of  equations
(\ref{eq:DCP}) and therefore an acceptable one for our interpolation
problem,   even if more optimal solutions may exist. Furthermore, as
we will see later in the examples, the method is very efficient and
defines very accurate solutions with a
small number of iterations. 

\end{itemize}
}
\subsection{Numerical integration: unitary methods}

Our construction  has produced the system of equations
(\ref{eq:coordinates}) for the solutions of the optimization
problem. It is crucial now to notice that these equations represent
the flow of the 
Hamiltonian system (\ref{eq:Hamilton}), the first one (the  
$x$-coordinate in our notation) being the coordinate expression of
von Neumann's equation, which was proved to be a Hamiltonian vector
field (see equation 
(\ref{HamVecField})) with respect to the canonical Lie-Poisson tensor
on $\mathfrak{u}^*(n)$. {\color{black}This Hamiltonian vector field represents the
infinitesimal generator of the coadjoint action 
of the unitary 
group $U(n)$} and therefore we know that its integral curves are
contained in the corresponding orbit, which is known to be a
symplectic submanifold of $\mathfrak{u}^*(n)$ (see
\cite{Abraham1978}).  In particular, if we 
consider an initial condition which is a pure state, the corresponding
symplectic orbit will be a curve on the set $\D^1(\mathcal H)$  and therefore 
diffeomorphic to a curve in  $P\mathcal{H}$.  {\color{black}Notice that, being a
unitary transformation, the evolution also defines isometries for the
scalar product  on $\u^{*}(n)$. 

If we are considering the interpolation of a set of points which do
not belong to the same unitary orbit, the situation is a bit
different. As the evolution we choose (with the vector field
associated to von Neumann equation) is always
unitary, any solution of equations (\ref{eq:coordinates}) must define a unitary
transformation. Hence, the minimal distance  of the trajectory 
to  the target point corresponds to the distance between the unitary orbit
containing the initial condition chosen and the target point.  Notice
that the algorithm introduced in the previous section makes sense also
in this case: the sequence of operators $K_{j}$ defined by equation
(\ref{eq:10}) steer the solutions towards the point in the unitary
orbit which minimizes the distance with respect to the target
point. We will discuss this point in detail in the following section
for a particular example.}

From both properties we know that, by implementing a numerical method
which preserves the symplectic structure and the distance, we will be
ensuring that trajectories  of equation (\ref{eq:coordinates}) will
remain on the corresponding orbit and therefore that  they will define
solutions of our interpolation problem. There exists several numerical
methods with those properties,  being one of the best known the
Gauss(-Legendre) Runge-Kutta (Gauss RK) implicit method (see
\cite{Dieci1994} and \cite{Hairer2006}), that we will be using in the
example of the next Section. The transition matrix of the integrator
defines thus a canonical transformation (since the integrator is
symplectic) which is also an isometry for the metric structure. 

\section{Example 1: the interpolation problem for a qubit}
\label{Section4}

As an application of our method, 
this section presents  the optimal control problem on the Lie algebra
$\u^*(2)$,   with the  function (\ref{eq:CostFunction}), $L$,  defined
on $\u^*(2)\times \u^*(2)\times \u^*(2)$.  Our goal is to determine a
quantum spline for a set of points $\{\rho_j\}\in
\mathcal{D}^1(\mathcal{H})$ and a set of times $\{t_j\}$.

We will use, for simplicity,  the identification of $\u^{*}(2)$ with the set of
Hermitian matrices $\rmi \u(2)$ and consider as  basis the set of
Hermitian Pauli spin matrices: 
$$
\sigma_1=\left(\begin{array}{cc} 1 & 0\\ 0 & 1\end{array}\right),\quad
\sigma_2=\left( \begin{array}{cc} 0 & 1 \\ 1& 0 \end{array}\right),\quad
\sigma_3=\left( \begin{array}{cc} 0 & -i \\ i & 0 \end{array}\right)\qquad\mbox{and}\quad
\sigma_4=\left( \begin{array}{cc} 1 & 0 \\ 0 & -1 \end{array}\right),
$$
that  obey the commutation relations ${\color{black}[\sigma_2,\sigma_3]_{\rmi
  \u(2)}}=2\sigma_4$, ${\color{black}[\sigma_4,\sigma_2]_{\rmi \u(2)}}=2\sigma_3$,
${\color{black}[\sigma_3,\sigma_4]_{\rmi \u(2)}}=2\sigma_2$. The structure constants
with respect to this basis are 
$$
c_{23}^4=c_{42}^3=c_{34}^2=-c_{32}^4=-c_{24}^3=-c_{43}^2=2\quad \mbox{and zero otherwise}.
$$

The elements $\rho\in \mathcal{D}^1(\mathcal{H})\subset \rmi\u(2)$ are written  as
	\begin{equation}
	\rho(t)=\frac 12 \sigma_1+\sum_{k=2}^4x_k(t)\sigma_k,
	\end{equation} 
while the elements of the form $H\in T_\rho\rmi\u(2)\sim\rmi\u(2)$ become
\begin{equation}
  H(t)=\sum_{k=1}^4y_k(t)\sigma_k.
\end{equation} 
With respect to these coordinates, the system (\ref{eq:coordinates}) turns out to be
\begin{equation}
  \label{eq:5}
\begin{array}{l}
\dot x_1(t)= 0\\[5pt]
\dot x_2(t)=2x_4(t)\,y_3(t)-2x_3(t)\,y_4(t)\\[5pt]
\dot x_3(t)=2x_2(t)\, y_4(t)- 2x_4(t)\,y_2(t)\\[5pt]
\dot x_4(t)=2x_3(t)\,y_2(t)-2x_2(t)\,y_3(t)\\[5pt]
\ddot y_1(t)= 0\\[5pt]
\ddot y_2(t)= k_j^2+2y_3(t)\dot y_4(t)-2y_4(t)\dot y_3(t)\\[5pt]
\ddot y_3(t)= k_j^3+2y_4(t)\dot y_2(t)-2y_2(t)\dot y_4(t)\\[5pt]
\ddot y_4(t)= k_j^4+2y_2(t)\dot y_3(t)-2y_3(t)\dot y_2(t).
\end{array}\smallskip
\end{equation}
Note that the definition of $K_j$ implies $k^1_j=0$, for all
$j=1,\dots,N$. The dynamics also implies that the curve $\rho$ in the
set $\D^1(\mathcal{H})$ of rank-one-projectors can be identified  with
a curve on the border of a 
sphere of radius $r=\frac 12$,   that is,  $x_2^2+x_3^2+x_4^2=\frac 14$.
{\color{black} Equations (\ref{eq:8}) imply that the value of $u(t)$ at the
  final point of each subinterval must vanish and this is imposed as a
  boundary condition. The only free parameter is the value of $H(0)$,
  which can be seen to be irrelevant for the algorithm, since the
  implicit algorithm adapts itself to it. Remember also
  that  $H(t)$
  is required to be continuous.}
\medskip

Consider the uniform partition of times $t_0=0<t_1=\frac 15 <t_2=\frac
25<t_3=\frac 35<t_4=\frac 45<t_5=1$   and the following set of
rank-one-projectors: 
\begin{equation}
\begin{array}{l}
\rho_0 = \frac 12\,\sigma_1+\frac 12\,\sigma_4 \\[5pt]
\rho_1 = \frac 12\,\sigma_1 + \frac 14\,\sigma_2+ \frac 14\,\sigma_3+ \frac {\sqrt{2}}4\,\sigma_4 \\[5pt]
\rho_2 = \frac 12\,\sigma_1+  \frac 38\,\sigma_2 + \frac {\sqrt{3}}8\,\sigma_3 + \frac 14\,\sigma_4 \\[5pt]
\rho_3 = \frac 12\,\sigma_1 +\frac 12\,\sigma_2 \\[5pt]
\rho_4 = \frac 12\,\sigma_1 +  \frac {\sqrt{3}}8\,\sigma_2 +  \frac 18\,\sigma_3 -\frac {\sqrt{3}}2\,\sigma_4 \\[5pt]
\rho_5 = \frac 12\,\sigma_1 +\frac 12\,\sigma_3.
\end{array}
\end{equation}
The problem consists in finding the optimal control function $u(t)$ and
the optimal state trajectory $(\rho(t),H(t))$
that minimizes the functional 
\begin{equation}
\label{cost1}
 J(u)=\int_0^1 \frac{1}{2}\|u\|^2\,\rmd t
 +\frac{1}{2\epsilon}\sum_{j=1}^5 \|\rho(t_j)-\rho_j\|^2 
\end{equation}
and satisfies the dynamical system
\begin{equation}
	\dot
        \rho(t)={\color{black}-i(H(t)\rho(t)-\rho(t)H(t))}\qquad\mbox{and}\qquad\displaystyle
        \dot H(t)=u(t), 
\end{equation}
subject to the initial conditions
\begin{equation}
	\rho(0)=\rho_0 \qquad  \mbox{and} \qquad H(0)=\sigma_4,
\end{equation}
where regularity and interpolation conditions are assumed according to
the dynamical interpolation problem $P$. 

{\color{black}We have implemented our algorithm described in Section
\ref{sec:impl-numer-integr} for different values of the number of
iterations and also different values of the parameter $\epsilon$.
Notice that both options are not independent since a larger value of
$\epsilon$ leads to a faster convergence for the algorithm.  
The number of iterations of the algorithm also affects the value of
the continuous part of the functional $J$ (equation \ref{cost1}): the
larger the number of iterations, the larger the value for the integral
part.  In the table below we present the values obtained for the
distance to the target points but also for the continuous part of the
functional. }Numerical integration was done with Wolfram's Mathematica, by using
the implicit unitary integrator included in the ``ImplicitRungeKutta"
library of the \textit{NDSolveUtilities} package. It is immediate to
verify that the purity of the state does not change at all
        because of the unitarity of the integrator (hence the points
        are always on the surface of the sphere) 

\begin{figure}[h]
\begin{center}	
\includegraphics[width=5cm]{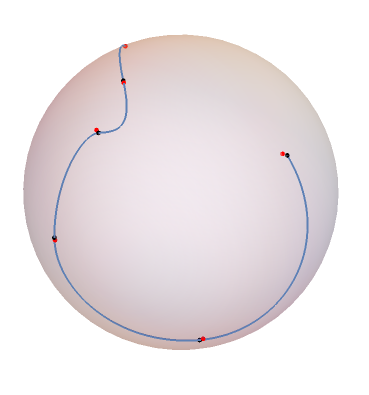}
	\hspace{2cm}
\includegraphics[width=5cm]{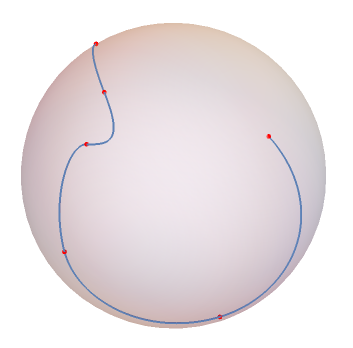}
\caption{{\color{black}Curve in $\D^1(\C^2)\equiv S^2$ for $\epsilon=0.005$ and 5
  iterations of the algorithm (left) and 50 iterations (right). Points
  $\rho_j$ for $j=0, \ldots, 5$ are represented in red. Points
  $\rho(t_j)$ are represented in black. We see how on the right, target
points and the points reached by the trajectory are
indistinguishable. In the figure on the left we can see  a
small discrepancy between red and black points.}}
 \label{figure}
 \end{center}
\end{figure}

The resulting curve can be found in  the figure \ref{figure} and it
produces a remarkably accurate result with suitable values for the
parameter $\epsilon$, as it can be seen in the tables  {\color{black}
  for $\epsilon=0.005$ and 5 iterations (left)  and for  $\epsilon=0.005$ and
50 iterations (right)}

  \begin{center}
  		\begin{tabular}{|c|c|c|c|}
			\hline 
			$t$&$\| \rho(t_j)-\rho_j\|$
                  & $J_{cont} $ & $J$\\[3pt] 
			\hline 
                  $0$&$0$  & $0$& $0$ \\[3pt]
                       			\hline 
                  $0.2$& $0.0101$&
                 $54.90$  & $54.91$\\[3pt]  
			\hline 
                  $0.4$& $0.0125$& 
                                                             $72.46$
                                & $72.47$\\[3pt]  
			\hline 
			$0.6$	& $0.0077$  &
                                                             $33.73$
                  & $33.73$\\[3pt]  
			\hline 
                  $0.8$&	$0.0128$ &
                                                                  $55.87$&
                  $ 55.88$\\[3pt]  
			\hline 
                  $1$	& $0.0178$ &
                                                                           $62.19$
                  & $62.22$\\[3pt]  
			\hline 
		\end{tabular} 
		\begin{tabular}{|c|c|c|c|}
			\hline 
			$t$&$\| \rho(t_j)-\rho_j\|$& $J_{cont} $ & $J$\\[3pt] 
			\hline 
                  $0$&$0$  & $0$&$0$\\[3pt]
                       			\hline 
                  $0.2$& $7.14\times 10^{-11}$&
                 $57.82$ & $57.82$\\[3pt]  
			\hline 
                  $0.4$& $7.16\times 10^{-11}$& 
                                                             $85.07$
                                                                 & $85.07$\\[3pt]  
			\hline 
			$0.6$	& $7.16\times 10^{-11}$  &
                                                             $46.12$
                  & $46.12$\\[3pt]  
			\hline 
                  $0.8$&	$7.16\times 10^{-11}$ &
                                                                       $
                                                         47.55$ & $47.55$\\[3pt]  
			\hline 
                  $1$	& $7.15\times 10^{-11}$ &
                                                                           $60.80$
                  & $60.80$\\[3pt]  
			\hline 
		\end{tabular} 

	\end{center}

	We verify that the distance to the target points can be made
        almost zero for a sufficiently large number of iterations. At each intermediate point $t_k$, we also include the values of the continuous part of the functional ($J_{cont}=\int_{t_{k-1}}^{t_k}\frac
        12 \| u\|^{2}dt$) computed over the subinterval and the total value of $J$.
        
	The computation time on a PC takes less than two seconds
        per iteration  for  the example presented.  We will see in
        next Section how the time increases significantly when
        considering a three-level case.

\section{Example 2: the interpolation problem for a qutrit}
\label{Section5}
As a second application of our method,  this section presents  the
case of a three-level systems (a qutrit). The formulation is entirely
analogous to the previous case but now the problem is formulated on
the space $\mathcal{D}(\C^{3})\subset \u^{*}(3)$. As we did in the
qubit case we will use the identification of $\u^{*}(3)$ with $\rmi
\u(3)$ and we will represent the density states as $3\times 3$
Hermitian matrices. As a basis, we will consider the set of Gell-Mann
matrices

$$
\lambda_{1}=
\left(
\begin{array}{ccc}
 0 & 1 & 0 \\
 1 & 0 & 0 \\
 0 & 0 & 0 \\
\end{array}
\right),
\qquad
\lambda_{2}=
\left(
\begin{array}{ccc}
 0 & -i & 0 \\
 i & 0 & 0 \\
 0 & 0 & 0 \\
\end{array}
\right),
\qquad
\lambda_{3}=
\left(
\begin{array}{ccc}
 1 & 0 & 0 \\
 0 & -1 & 0 \\
 0 & 0 & 0 \\
\end{array}
\right)
$$

$$
\lambda_{4}=
\left(
\begin{array}{ccc}
 0 & 0 & 1 \\
 0 & 0 & 0 \\
 1 & 0 & 0 \\
\end{array}
\right),
\qquad
\lambda_{5}=
\left(
\begin{array}{ccc}
 0 & 0 & -i \\
 0 & 0 & 0 \\
 i & 0 & 0 \\
\end{array}
\right),
\qquad
\lambda_{6}=
\left(
\begin{array}{ccc}
 0 & 0 & 0 \\
 0 & 0 & 1 \\
 0 & 1 & 0 \\
\end{array}
\right)
$$

$$
\lambda_{7}=
\left(
\begin{array}{ccc}
 0 & 0 & 0 \\
 0 & 0 & -i \\
 0 & i & 0 \\
\end{array}
\right),
\qquad
\lambda_{8}=\frac{1}{\sqrt{3}}
\left(
\begin{array}{ccc}
  1& 0 & 0 \\
 0 & 1 & 0 \\
 0 & 0 & -2 \\
\end{array}
\right),
\qquad
\lambda_{9}=
\sqrt{\frac{2}{3}}
 \left(
\begin{array}{ccc}
 1 & 0 & 0 \\
 0 & 1 & 0 \\
 0 & 0 & 1 \\
\end{array}
\right).
$$
 The structure constants with respect to this basis are 
$$
c_{12}^3=-c_{21}^{3}=2;\quad 
c_{45}^8=c_{67}^8=-c_{54}^{8}=-c_{76}^{8}=\sqrt{3};
$$
$$
c_{14}^7=-c_{41}^7=-c_{15}^{6}=c_{51}^{6}=c_{24}^{6}=-c_{42}^{6}=c_{25}^{7}=-c_{52}^{7}=c_{34}^{5}=-c_{43}^{5}=c_{63}^{7}=-c_{36}^{7}=1
$$
and zero otherwise. The elements $\rho\in
\mathcal{D}(\mathcal{H})\subset \rmi\u(3)$ are written  as 
	\begin{equation}
	\rho(t)=\frac 12 \lambda_9+\sum_{k=1}^8x_k(t)\lambda_k,
	\end{equation} 
while the elements of the form $H\in T_\rho\rmi\u(3)\sim\rmi\u(3)$ become
\begin{equation}
  H(t)=\sum_{k=1}^9y_k(t)\lambda_k.
\end{equation} 
With respect to these coordinates, we consider now  the problem of
finding the optimal control function $u(t)$ and  the optimal state
trajectory $(\rho(t),H(t))$ that minimizes the functional 
\begin{equation}
\label{cost}
 J(u)=\int_0^1 \frac{1}{2}\|u\|^2\,\rmd t
 +\frac{1}{2\epsilon}\sum_{j=1}^N \|\rho(t_j)-\rho_j\|^2 
\end{equation}
and satisfies the dynamical system
\begin{equation}
	\dot
        \rho(t)=-i(H(t)\rho(t)-\rho(t)H(t))\qquad\mbox{and}\qquad\displaystyle
        \dot H(t)=u(t), 
\end{equation}
subject to the initial conditions
\begin{equation}
	\rho(0)=\rho_0 \qquad  \mbox{and} \qquad H(0)=\lambda_{9},
\end{equation}
where regularity and interpolation conditions are assumed according to
the dynamical interpolation problem $P$.  As we saw above, we have to
solve the  system (\ref{eq:coordinates}).
Again, equations (\ref{eq:8}) imply that the value of $u(t)$ at the
  final point of each subinterval must vanish and this is imposed as a
  boundary condition. The value of $H(0)$
  can also be seen to be irrelevant for the algorithm, as it happens
  in the case of two levels.
\subsection{First example: points contained in a unitary orbit}

Let us consider the following set of points, for times
$t_0=0<t_1=\frac 16 <t_2=\frac 13<t_3=\frac 12<t_4=\frac 23<t_5=\frac
56 < t_{6}=1$:
$$
\rho_{0}=
\left(
\begin{array}{ccc}
 \frac 13 & 0 & 0 \\
 0 & \frac 23 & 0 \\
 0 & 0 &  0\\
\end{array}
\right),
$$
$$
\rho_{1}=
\left(
\begin{array}{ccc}
 0.436919& -0.0234205-0.187994 i & 0.109777\, +0.158205 i \\
 -0.0234205+0.187994 i & 0.442465& 0.0387764\, +0.0518969 i \\
 0.109777\, -0.158205 i & 0.0387764\, -0.0518969 i & 0.120616\\
\end{array}
\right),
$$
$$
\rho_{2}=
\left(
\begin{array}{ccc}
 0.25208\,  & 0.0710467\, -0.0594233 i & -0.178472+0.143899 i \\
 0.0710467\, +0.0594233 i & 0.358968 & 0.0437509\, -0.207081 i \\
 -0.178472-0.143899 i & 0.0437509\, +0.207081 i & 0.388953 \\
\end{array}
\right),
$$
$$
\rho_{3}=
\left(
\begin{array}{ccc}
 0.510032 & 0.0421306\, -0.160051 i & -0.158675+0.163612 i \\
 0.0421306\, +0.160051 i & 0.268756 & 0.0865724\, +0.0172119 i \\
 -0.158675-0.163612 i & 0.0865724\, -0.0172119 i & 0.221213\\
\end{array}
\right),
$$
$$
\rho_{4}=
\left(
\begin{array}{ccc}
 0.145442 & 0.0762356\, -0.126603 i & -0.0740697+0.211438 i \\
 0.0762356\, +0.126603 i & 0.450398& -0.107868-0.0207668 i \\
 -0.0740697-0.211438 i & -0.107868+0.0207668 i & 0.40416\\
\end{array}
\right),
$$
$$
\rho_{5}=
\left(
\begin{array}{ccc}
 0.294447& 0.1995\, -0.0726447 i & -0.0696641+0.219691 i \\
 0.1995\, +0.0726447 i & 0.301392& -0.0818337-0.0494491 i \\
 -0.0696641-0.219691 i & -0.0818337+0.0494491 i & 0.40416 \\
\end{array}
\right),
$$

$$
\rho_{6}=
\left(
\begin{array}{ccc}
 0.0638338 & -0.040794+0.00995446 i & 0.00664582\, -0.133158 i \\
 -0.040794-0.00995446 i & 0.522085 & 0.152599\, -0.104391 i \\
 0.00664582\, +0.133158 i & 0.152599\, +0.104391 i & 0.414082\\
\end{array}
\right).
$$
All seven points belong to a unitary orbit of the set of mixed states in
$\mathcal{D}(\C^{3})$. 

We have considered again our algorithm for this case in order to
verify how does it work for the case of mixed states.  We have considered a value
for $\epsilon=0.001$ and 200 iterations of our algorithm.  The
resulting distances of the points obtained and the values of the
functional $J$ and its continuous part $J_{cont}$ are the following:
\vspace*{0,2cm}

\begin{center}
\begin{tabular}{|c|c|c|c|}
\hline 
$t$&$\| \rho(t_j)-\rho_j\|$& $J_{cont}$ & $J$\\[3pt] 
\hline 
  $0$&$0$ & 0& 0\\[3pt]
\hline 
 $\frac 16$& $4.87\times 10^{-10}$&$1.82\times 10^{-16}$& $3.01\times 10^{-16}$  \\[3pt]  
\hline 
  $\frac 13$& $6.7\times 10^{-4}$& $845.53$ & $845.53$\\[3pt]  
\hline 
$\frac 12$	& $9.01\times 10^{-7}$& $607.02$& $607.02$\\[3pt]  
\hline 
$\frac 23$&	$9.51\times 10^{-9}$ &$77.55$& $77.55$\\[3pt]  
\hline 
$\frac 56$	& $5.64\times 10^{-7}$ &$298.54$ &$298.54$ \\[3pt]  
\hline
$1$	& $8.9\times 10^{-6}$&$282.83$ &$282.83$ \\[3pt]  
\hline
\end{tabular} 
\end{center}

\vspace*{0,2cm}
Computation times are now longer, up to 5-6 seconds per iteration in
certain cases.
We conclude thus that the algorithm is also very efficient in the case
of mixed states, although smaller values of the parameter $\epsilon$
must be considered. Notice that this example constitutes an important
generalization with respect to the approach presented in
\cite{Brody2012} which makes sense only for pure states.

\subsection{Points in different unitary orbits}

        Furthermore, the richness of the structure of unitary orbits
 of the set $\mathcal{D}(\C^{3})$ allows us to consider different
 situations. For instance, we can consider a case where the target
 points do not belong to the same unitary orbit and check the behavior
 of the algorithm. If we consider as initial point
 $$
 \rho_{0b}=
 \left(
\begin{array}{ccc}
 \frac 13-0.001 & 0 & 0 \\
 0 & \frac 23+0.001 & 0 \\
 0 & 0 &  0\\
\end{array}
\right),
 $$
and seek for a curve joining $\rho_{0b}$, $\rho_{1}$ and $\rho_{2}$ by
using equations  (\ref{eq:coordinates}) we obtain points which are
contained in the unitary orbit of $\rho_{0b}$. Hence, as $\rho_{1}$
and $\rho_{2}$ lie in a different orbit, the curve can not reach
them. Nonetheless, the algorithm stabilizes (after 100 iterations) at
points
$$
\rho_{1b}=
\left(
\begin{array}{ccc}
 0.437204 & -0.0233546-0.18881 i & 0.109899\, +0.158095 i \\
 -0.0233546+0.18881 i & 0.442331 & 0.0383441\, +0.0521308 i \\
 0.109899\, -0.158095 i & 0.0383441\, -0.0521308 i & 0.120465 \\
\end{array}
\right)
$$
and
$$
\rho_{2b}=
\left(
\begin{array}{ccc}
 0.251872 & 0.070592\, -0.0598828 i & -0.178803+0.143833 i \\
 0.070592\, +0.0598828 i & 0.358735 & 0.0435776\, -0.207622 i \\
 -0.178803-0.143833 i & 0.0435776\, +0.207622 i & 0.389393 \\
\end{array}
\right)
$$
which are at a distance of 0.001 of $\rho_{1}$ and $\rho_{2}$,
respectively. These are the closest possible points since the unitary
transformation must preserve the distance between both unitary orbits
which is 0.001 at the initial points ($\rho_{0}$ and $\rho_{0b}$).

We conclude thus that our method provides an efficient numerical solution for the
interpolation problem of general (pure or mixed) quantum states and 
generates a trajectory contained in the closest unitary orbit to them.

\section{Conclusions and outlook}
\label{Conclusions}
{\color{black}
In this paper we have generalized the notion of quantum spline
introduced in  \cite{Brody2012} to the case of general quantum states,
pure or mixed.  } Brody \emph{et al.} formulated a variational problem
on a projective space defined in Section \ref{Section1}. They
considered a globally  defined continuous curve on the Lie group
$U(n)$ and the corresponding set of continuous  variations, and study
the set of extremals of the functional defined in equation
(\ref{eq:JproblemA}), which combines the curve on $U(n)$ and its
action on the complex projective space.  From this, they 
obtain a set of differential equations whose solutions correspond to
Riemannian cubic polynomials on 
 the subintervals $]t_{j-1}, t_{j}[$ and a set of boundary conditions on
 the times $\{t_j\}$. Finding a global solution is a difficult problem
 and they also provide a numerical algorithm illustrated by a
 particular example in  the case of a two level system.

{ \color{black}
Our definition reconsiders the problem and writes it on the complete
set of quantum states (pure and mixed), modeled as a submanifold $\mathcal{D}$ of
the dual space  $\u^{*}(n)$. Considering (as in \cite{Brody2012} ) the
case of unitary dynamics, allows us to forget about $\mathcal{D}$ and
define the problem globally on the linear space $\u^{*}(n)$. This
simplifies the problem in a remarkable way. Furthermore, we consider
a different type of optimization problem and choose a local
formulation on the different subintervals of the time domain which
does not impose the differentiability of the time dependence of the
Hamiltonian. This choice, which is common, for instance,  in quantum
control solutions based on pulses, simplifies further the problem from
the mathematical point of view, and transform it in a chain of
Bolza-type problems.  We consider then a Hamiltonian formulation of
the corresponding control problem based on Pontryagin 
Maximum Principle, which allows us to consider in a natural way
symplectic integrators associated to the geometrical formulation of
Quantum Mechanics. Despite the much simpler formulation, the resulting
system of differential equations is still difficult to solve
exactly. We have introduced then an iterative algorithm to determine
good solutions of the problem which works very efficiently for 
pure states and for  mixed states, even for problems where the
points to be interpolated do not belong to the same unitary orbit. Our
algorithm allows us to define the closest unitary orbit in an
efficient way. 

Another advantage of our formalism is that it admits further
generalizations in a simple way. Indeed, we can consider an analogous
formulation where unitary dynamics is replaced by more general master
equations, as for instance the Lindblad-Kossakowski equation. This
Markovian generalization may impose some extra constraints on the set
of possible times and points to be interpolated, but the formulation
of the problem makes perfect sense. The conclusions of that analysis,
which is being done now, will be presented in a future paper.

}

\section*{Acknowledgments}

The work of L. Abrunheiro was supported by Portuguese funds through the \emph{Center for Research and Development in Mathematics and Applications} (CIDMA)  and the \emph{Portuguese Foundation for Science and Technology}
(``FCT --Funda\c{c}\~ao para a Ci\^encia e a Tecnologia''), within project UID/MAT/04106/2013. The work of M. Camarinha and P. Santos was partially supported by the Centre for Mathematics of the University of
Coimbra -- UID/MAT/00324/2013, funded by the Portuguese Government through FCT/MEC and co-funded by the European Regional Development Fund through the Partnership Agreement PT2020. The work of J.
Clemente-Gallardo and J. C. Cuch\'{\i} was partially covered by MICINN  grants FIS2013-46159-C3-2-P and MTM2012-33575 and by DGA Grant 2016-24/1.




\end{document}